# Optical surface edge Bloch modes: low-loss subwavelength-scale 2D light localization

Shu-Yu Su[*] and Tomoyuki Yoshie

*Electrical and Computer Engineering, Fitzpatrick Institute for Photonics, Duke University, Durham, NC 27708*
*Corresponding author: shuyu.su@duke.edu

*An optical surface edge mode is an optical state evanescently bound on an edge of intersecting photonic crystal lattice terminations. Two intersecting surfaces of a 3D photonic crystal form an edge of intersecting lattice terminations. Low-loss subwavelength-scale edge modes can appear on an <010> edge of a dielectric woodpile within a complete bandgap. The mode area is as small as 0.066 squared half-in-vacuum-wavelengths. The edge mode has field maxima in vacuum near the termination surface, like surface plasmon modes. This edge mode would provide new opportunities of low-loss light localization in a sub-diffraction-limit space without the use of metal.*

An optical surface Bloch mode [1-7] is an optical state evanescently bound at a surface on a periodic structure, and the mode frequency is located within a photonic bandgap (PBG). Tamm [8] studied electronic surface Bloch modes. Yeh, Yariv and Hong [1] pioneered the study of optical surface Bloch modes in one-dimensional (1D) periodic dielectric media. The optical surface Bloch mode is an intriguing subject in optical physics and condensed matter physics and is potentially useful for nanophotonic applications such as optical sensing and cavity quantum electromagnetic dynamics (QED) experiments, as the mode is localized at the surface of a photonic crystal within subwavelength scales.

A finite-size photonic crystal has multiple crystal terminations (representatively surface planes) unless it has cyclically periodic conditions—Born-von Karman boundary conditions. Figure 1 shows a finite-size woodpile three-dimensional (3D) photonic crystal terminated on both (100) and (001) surface planes. Upper and right panels in Fig. 1 show a (100) surface plane mode on an infinite (100) surface and a (001) surface plane mode on an infinite (001) surface, respectively. The <100>, <010>, and <001> crystal directions are parallel to the x, y, and z axes. For 3D photonic crystals, an intersection of two (three) non-parallel surface planes forms an edge (a vertex). Optical localization at surface intersections is important as it provides new opportunities of low-loss light localization in subwavelength scales without the use of metal. In our work, we consider woodpile dielectric photonic crystals, as they are representative 3D photonic crystals with a complete bandgap and can be fabricated by two-directional etching method [9,10] and by layer-by-layer method [5,11]. Here we report novel localized modes at an edge on a woodpile 3D photonic crystal embedded in vacuum.

Three-dimensional plane-wave expansion method [12] is used to analyze the propagation modes bound on faces and edges on finite-size woodpile dielectric photonic crystals. We made sure that the surface modes decayed sufficiently in vacuum along the transverse directions (i.e. the directions orthogonal to the <010> propagation direction) so that they did not have the coupling through computational boundary walls in the transverse directions. A discretized spatial permittivity function representing a supercell is Fourier transformed for the plane wave expansion method calculation. The resolution—the number of pixels in one lattice constant representing the spatial permittivity—is set to 22 for Figs. 1-3 and 36 for Fig. 4. The lattice constants of the woodpile photonic crystal are $a_x = a_y = a$ and $a_z = 1.2a$. The dielectric rods have refractive index of 3.4, and their height h and width w are both 0.3a. The surface termination parameter is defined as $\{t_i = (i_0-i)/a_i: 0 \leq t < 1\}$ (i=x,y,z); see Fig. 1 and Fig. 3.

A finite-size woodpile photonic crystal supports transverse modes and thus has more surface plane Bloch modes than a semi-infinite-size woodpile. The standing transverse planar modes are formed because a surface plane mode is terminated at two ends and reflected transversely between the two ends of the finite-size woodpile. Figure 2 (a) shows dispersion curves of surface Bloch modes propagating along the <010> direction for $4\times\infty\times4$ unit cells. Surface modes are bound on one face, on an edge, or on multiple faces. Figure 2 (b) shows modes appearing on the top and the right surfaces including the mode on the edge; see Fig. 1 for finding "top" and "right" surfaces. We find four surface modes; two on the right (100) surface plane, one on the top (001) surface plane, and one on the (100)-(001) intersection, i.e. the <010> edge. Figure 2 (c) shows the surface edge Bloch mode on the <010> edge.

These surface modes can be tuned by the surface termination parameters ($t_x, t_z$). We show surface Bloch modes bound only on a particular combination of two intersecting planes in Fig. 3; as we did in Fig. 2 (b), we removed modes bound on left and bottom surfaces on a finite-size woodpile. Figure 3 includes (100) right surface plane modes, (001) top surface plane modes, (100)-and-(001) surface modes, and <010> edge modes. On the other hand, microcavity modes at a vertex on a woodpile have not been found from our analysis so far.

One distinct result is the formation of ultralow-loss, subwavelength-scale edge modes. For $4\times\infty\times4$ unit cells, the radiation quality factor (Q factor) of this waveguide edge mode is $3\times10^7$ and the mode area A is 0.066 $(\lambda_0/2)^2$ which is in subwavelength-scales. The wavelength in vacuum is $\lambda_0$. Optical field of this edge mode is localized on the lattice termination edge and the field maxima are in vacuum near the termination surface, as seen in Fig. 2(c); therefore, we express the mode area in unit of $(\lambda_0/2)^2$ in lieu of $(\lambda_0/2n)^2$. The waveguide Q factor is obtained by calculating $\omega U/P$ where $\omega$, U, and P are the angular frequency, the stored energy, and the power loss along the transverse directions, i.e. out-of-propagation directions, so it represents the radiation Q factor. The mode area is calculated from the relation $A = \int U(x,y,z)dV / (a \max[U(x,y,z)])$ where U(x,y,z) is the stored electromagnetic energy density.

The field decays exponentially in vacuum, and the field oscillation decays exponentially in the photonic crystal by the nature of decaying Bloch waves in a complete bandgap. For non-surface modes in an isotropic dielectric waveguide, the field at maximum filed points is oscillatory like cosine function, and the waveguide requires at least ~$\lambda_0/2n$ length for optical confinement. This prevents us from building true subwavelength-scale waveguides and microcavities from isotropic dielectrics in linear optics regime. However, for surface Bloch modes, it is possible to avoid such oscillatory function at field maxima and remove a $\lambda_0/2n$ space. The surface edge Bloch modes are similar to surface plasmon edge states [13] in that both are evanescently localized at an edge of a structure. However, for isotropic dielectrics, evanescently decaying fields are possible via the introduction of Bloch waves.

The realization of surface edge mode structure would require high-precision fabrication. One major fabrication irregularity would be fluctuations of lattice termination which impact the dispersion. Local size fluctuations would also increase the scattering loss. We think that we can fabricate precise structures for long wavelength experiments such as RF experiments.

Other intriguing fact is that there are many transverse surface modes supported over the complete bandgap for finite-size woodpiles and the field decay is gradual. In-crystal optical modes (waveguides and microcavities) as well as photonic density of states would be influenced by surface modes. For reported 3D photonic crystals with a complete bandgap, the number of periods is still small. For example, lasing action was demonstrated from the microcavity (Q ~ 38,500) in a woodpile with a size of ~11×11×6 unit cells [11]—the shortest surface-to-cavity distance is only 3 periods in the z direction. For certain termination combinations in Fig. 3, there is a large frequency range in which no surface modes appear. Therefore, it may be wise to choose certain photonic crystal surfaces.

We analyzed co-directional coupling of two parallel (100) surface plane modes of the same type—one on a right infinite (100) surface plane with $t_x = 0$ and the other on a left infinite (100) surface plane with the same type of termination—to show slow evanescent field decay. Measuring a mode splitting frequency of two coupled (100) surface modes separated by a woodpile is a simple way to measure the coupling coefficient. There are symmetric and asymmetric supermodes, and the mode frequencies are denoted as $f_s$ and $f_a$. We introduce the coupling coefficient $\kappa = \pi |f_s-f_a| n_{ga}/c$ [14] where $n_{ga}$ is the average of group indices of two supermodes and c is the speed of light. Figure 4 shows the coupling coefficient κ as a function of the separation distance $N_x$. The coupling slowly decreases as the value $N_x$ increases. The analysis results suggest that a large photonic crystal or surface designing would help to suppress the coupling of the external continuum states to the in-crystal states via surface modes.

In conclusions, we analyzed surface Bloch modes of finite-size woodpile 3D photonic crystals. Edge modes are localized on a <010> edge for certain combinations of termination surfaces. The mode area is as small as 0.066 $(\lambda_0/2)^2$, and the waveguide Q factor is high thanks to dielectric optical confinement assisted by a complete photonic bandgap. The field maxima occur in vacuum near the dielectric-vacuum interface. Two-dimensional light localization at edges provides new opportunities in subwavelength-scale light localization in isotropic dielectric materials. We find many transverse surface Bloch modes over a complete photonic bandgap of a finite-size woodpile. As the evanescent field slowly decreases, a large photonic crystal or careful surface designing would be useful for observing distinct characteristics of optical field in a complete bandgap.

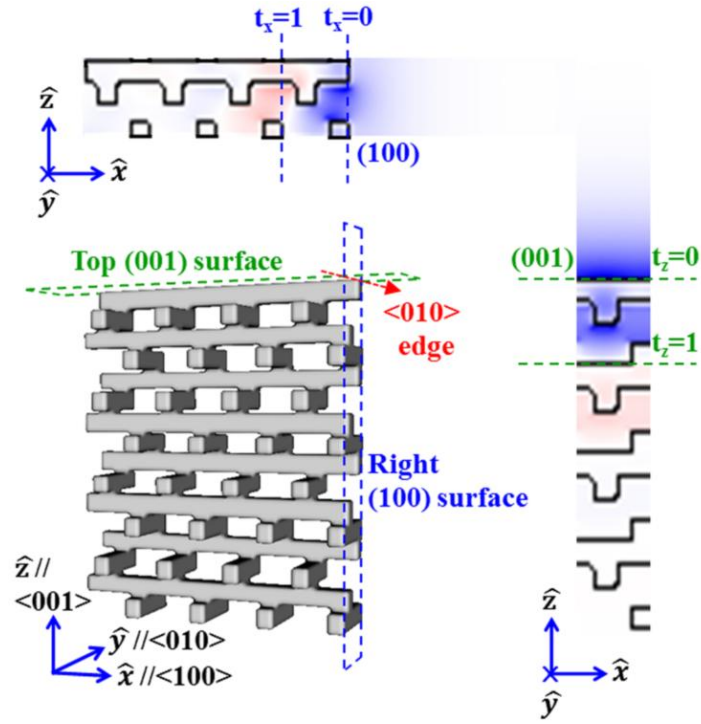

Fig. 1. A finite-size woodpile 3D photonic crystal and two types of surface plane Bloch modes on a semi-infinite-size woodpile. The finite-size structure shows 4 × 1 × 4 unit cells; one unit cell corresponds to two primitive cells [10]. An intersection of (100) and (001) faces forms a <010> edge. The field profiles of (100) and (001) surface plane modes are shown by using the electric field component that is perpendicular to the termination surface.

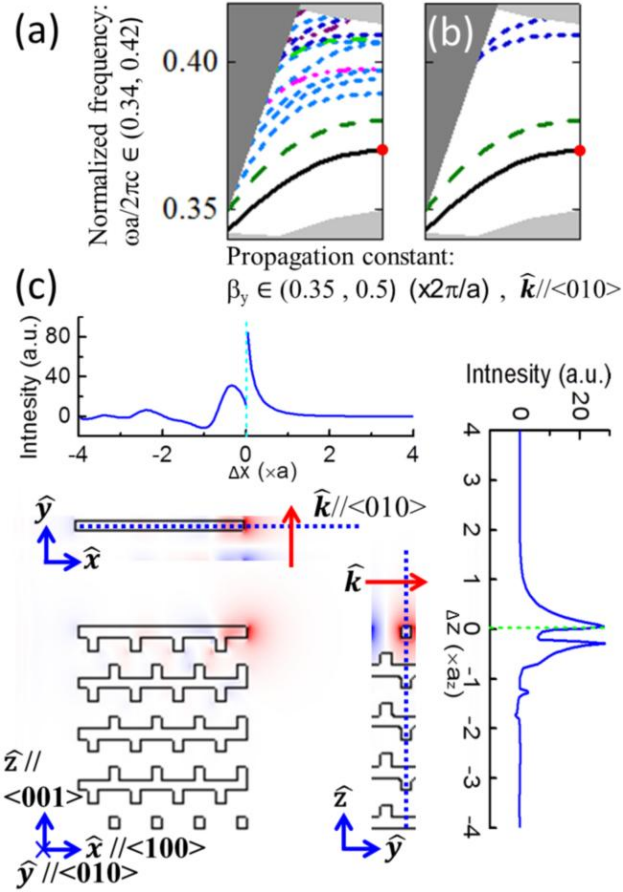

Fig. 2. Surface Bloch modes propagating along the <010> (y) direction on a finite-size woodpile and an edge mode: (a) Dispersion relations of surface modes on a woodpile of 4 × ∞ × 4 unit cells. (b) Dispersion relations of surface modes appearing only on two intersecting right (100) and top (001) surfaces with $(t_x, t_z) = (0, 0)$. The rest of modes seen in (a) appear on the other two planes, i.e. the left (100) and the bottom (001) surfaces. Solid black, dashed green, and dotted blue curves are dispersion functions for <010> edge modes, (001) plane modes, and (100) plane modes, respectively. (c) The mode profile ($E_x$ component) of the <010> edge mode at $\beta_y = \pi/a$, shown by red points in (a) and (b), viewed from different directions. Red arrows show the propagation direction **k**. Upper panel in (c) shows the field intensity versus $\Delta x$, the distance from the right (100) surface along the horizontal dotted blue line; right panel shows the field intensity versus $\Delta z$, the distance from the top (001) surface along the vertical dotted blue line, The edge mode has field maxima in vacuum near the termination surface.

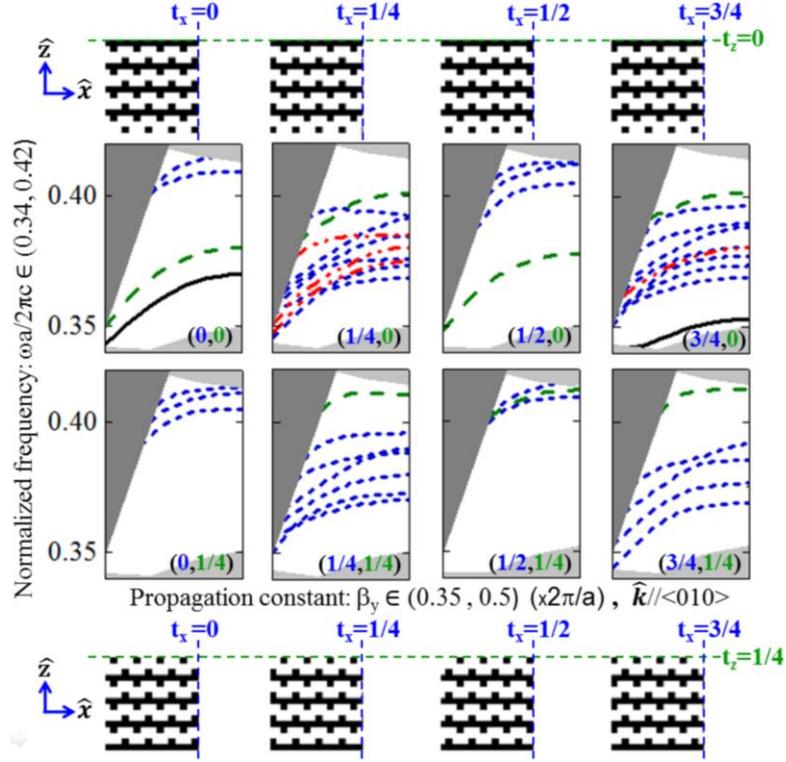

Fig. 3. Dispersion relations of surface Bloch modes versus lattice termination combination ($t_x$, $t_z$). Modes on a <010> edge, on a top (001) face, on a right (100) face, and on both top (100) and right (001) faces, shown by solid black, dashed green, dotted blue, and dash-dot-dot red curves, respectively. The white regions are mode gap regions within a complete bandgap projected along the propagation direction **k**. We did not show the cases of $t_z$ = 1/2 and 3/4 as the structures are equivalent for ($t_x$, $t_z$) and ($t_x$+1/2, $t_z$+1/2).

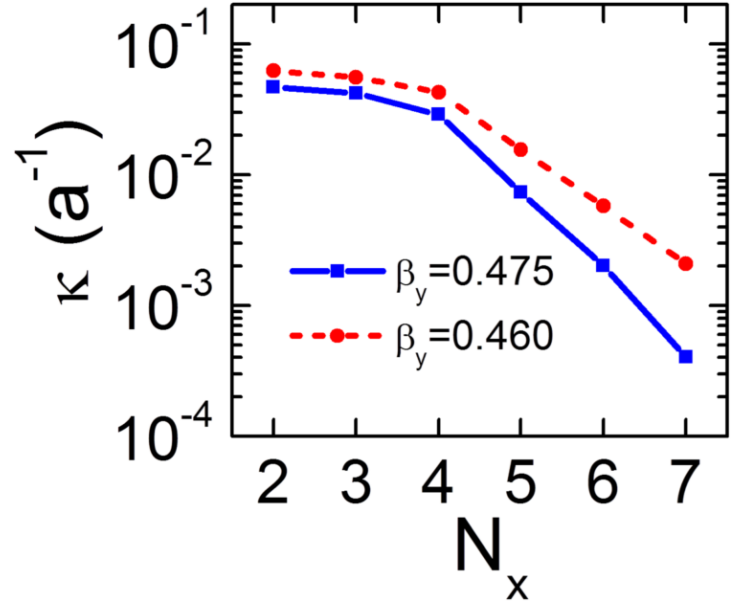

Fig. 4. The coupling coefficient κ versus $N_x$, i.e. the separation distance in the <100> direction in a unit of unit cell number, for two parallel (100) surface plane modes of the same type ($t_x$ = 0). The coupling coefficient is analyzed at appropriate propagation constants $β_y$ in order to avoid the coupling with other (100) surface modes.